\documentclass[jkps,preprint,fleqn,showpacs,showkeys]{revtex4}
\usepackage{graphicx}
\usepackage{amssymb}
\usepackage{amsmath}
\usepackage{bm}

\usepackage{amsfonts,amssymb,amsmath}
\usepackage{graphicx}

\usepackage{graphics,hyperref}
\usepackage{amssymb}
\usepackage{amsmath}
\usepackage{color}
\usepackage{bm}

\begin{document}
\setcounter{page}{0}
\title[]{A Brief Overview of Bipartite and Multipartite Entanglement Measures}
\author{Saeed \surname{Haddadi}}
\email{haddadi@physicist.net}
\author{Mohammad \surname{Bohloul}}

\affiliation{Department of Physics, Payame Noor University, P.O. Box 19395-3697, Tehran, Iran.}

\date[]{Received 7 May 2018}

\begin{abstract}
Measuring entanglement is a demanding task in the field of quantum computation and quantum information theory. Recently, some authors experimentally demonstrated an embedding quantum simulator, using it to efficiently measure two-qubit entanglement. Here, we are reviewing some measures of entanglement which are used for pure and mixed states. Furthermore, we have reported the efficient bipartite and multipartite entanglement measures.
\end{abstract}

\pacs{03.65.Ud, 03.67.Mn, 03.67.-a}

\keywords{Measures of entanglement; Efficient measures; Entanglement properties.}

\maketitle

\section{Introduction}
\label{intro}
Quantum entanglement plays a key ingredient of quantum information theory, such as quantum communication and quantum computation \cite{Schrodinger01,Einstein02}.  Quantum entanglement represents one of the most remarkable features of quantum systems that has no classical counterpart \cite{Guhne04}. Entanglement is an important quantum resource in both quantum computation and communication \cite{Sun5}. Entangled states indicate a variety of non-local quantum correlation subsystems, which have many applications in quantum data, including quantum teleportation, quantum cryptography \cite{Ekert}, quantum dense coding and quantum computing \cite{Sheng7,Zheng6}. The study of entanglement of such states is essential and recently has been a field of severe research \cite{Arnaud03}.
Candia \textit{et al.} \cite{Candia} proposed a protocol for the efficient measurement of multipartite entanglement with embedding quantum simulators \cite{Chen111}, but we have to focus on the meaning of the term efficient. It depends if we want to qualify or quantify the entanglement. For example, for pure states, we have some methods in quantum information theory that describe a certain type or class of entanglement, under stochastic local operations with classical communication (SLOCC) \cite{Dur05,Liang06}. Also by using invariant theory, some polynomial can help us to distinguish different classes of entanglement \cite{Messina13}. We can also study the entanglement of pure state by associating each quantum state, which is a geometrical singularity, that can help to have a piece of information on the entanglement of the state. Cavalcanti \textit{et al.} \cite{Cavalcanti} have demonstrated how the geometry of the set of unentangled states can be related to singular behavior in physical phenomena. Their results have proposed an explanation by interpreting the non-analyticities exhibited by entanglement as a consequence of geometric singularities \cite{Cavalcanti}. There are also some numerical measurements of entanglement that can help us to quantify entanglement, such as concurrence \cite{Hill07,Wootters08,sheng2000}, negativity and logarithmic negativity \cite{Vidal09,Plenio}, Von Neumann entropy, etc \cite{Nielsen10}. In this paper, we are looking for the efficient bipartite and multipartite entanglement measures. This paper is organized as follows: in Section \ref{sec:2} we provide bipartite systems and then we introduce multipartite systems in the Section \ref{sec:3}. Finally, Section \ref{sec:4} is dedicated to discussion and conclusions.

\section{Bipartite systems}
\label{sec:2}
When a system consists of two subsystems we say it is a bipartite system. The Hilbert space of the composite system is a tensor product of Hilbert space that describes Alice$^{,}$s system and the Hilbert space that describes Bob$^{,}$s system. If we define these as $H_{A}$ and $H_{B}$, respectively, then the Hilbert space of the composite system is $H=H_{A}\otimes H_{B}$. But not all states $|\psi\rangle \in H_{A}\otimes H_{B}$ are entangled. When two systems are entangled, the state of each composite system can only be described with reference to the other state. If $|\psi\rangle \in H_{A}$ and $|\phi\rangle \in H_{B}$ and $|\xi\rangle=|\psi\rangle \otimes |\phi\rangle$, then $|\xi\rangle$ is a product state or separable.
For bipartite systems, various measures of entanglement have been proposed \cite{Hill07,Wootters08,Vidal09,Nielsen10,Bennett11,Plenio12,Zhou1000}. Entanglement measure (EM) quantifies how much entanglement is in a bipartite or multipartite systems. Properly it is any non-negative real function of a state which cannot increment under local operations and classical communication (LOCC) \cite{Chitambar}, and is zero for product state or separable state. An LOCC operation is an element of the class LOCC, which contains all local quantum operations and classical communication. In other words, an LOCC operation would be doing a local quantum operation or doing some classical communication \cite{Nielsen10}. A good and rigorous definition of LOCC is not so easy, see \cite{Chitambar} for a nice review on LOCC. One of general applications of abstract entanglement measures (EMs) is to show that certain task cannot be obtained by means of LOCC \cite{Nielsen10,Chitambar}. One does it by showing that if the task could be done, then some of EMs would increment. EMs are also classified based on their properties, e.g. additivity, convexity and continuity. This approach to EMs is known as axiomatic approach \cite{Plenio13}. An EM for a bipartite system is a state functional that vanishes on separable states and does not increase under separable operations.

\section{Multipartite systems}
\label{sec:3}

The issue of defining multipartite entanglement is more difficult and there is no unique definition \cite{Bruss}. Hence, a good definition of multipartite entanglement should hinge upon some statistical information concerning the system. Some recent works have focused on clarifying concept of multipartite entanglement \cite{Scott27,Roscilde28}. Facchi \textit{et al.} began the study of multipartite entanglement from \cite{Facchi100} where multipartite entanglement has been characterized in terms of a probability distribution, namely the distribution of bipartite entanglement over all the possible bipartition. This is motivated by the fact that, due to the presence of many possible bipartitions, the characterization of multipartite entanglement resembles a complex system. This is taken into account since entanglement is not measured by a single number but by a whole probability distribution (thus all its moments). Facchi \textit{et al.} have recasted the characterization of multipartite entanglement in terms of a fictitious statistical mechanical problem \cite{Facchi101,Facchi102}, a technique that is often used in optimization problems. In this statistical mechanical approach, the hamiltonian is a suitable function that represents some features of multipartite entanglement (like one of the above moments or a combination of them), the configuration space is the set of all pure states, and the temperature is a parameter that fixes isoentangled (with respect to the features specified in the hamiltonian) subspaces \cite{Facchi101}. Correlation functions, high temperature expansions, and other techniques from statistical mechanics are used to characterize multipartite entangled states. This approach can also be seen as defining ensembles of random isoentangled states \cite{Facchi101}. The statistical mechanical approach has also been applied to gaussian states, to the characterization of bipartite entanglement of large systems, where phase transitions are identified and phases represent random states with support on subsets with different entanglement features \cite{Facchi103}, and to mixed states where now it is not entanglement to be characterized but local purities of the parties \cite{Pasquale}. In this framework, a state is maximally entangled if it is maximally entangled according to all bipartition. This condition is highly nontrivial, however, generically impossible for qubit systems or gaussian states except in the case of small number of constituents: different bipartitions interfere similarly to how local interactions do in frustrated systems \cite{Marzolino21,Facchi104}. Therefore, the problem of maximizing multipartite entanglement is nontrivial for two aspects: one is the complex nature of its characterization that jointly involves many quantities (the moments of the above distributions that all together carry the same information of the distribution itself), the second is the possible frustration in the optimization procedure due to the interference among bipartitions.

The study of multipartite entanglement has attracted much attention in the last years \cite{Gerke,LiLi,Zhang2016,akhound2016,Assadi,Pezz15}. From the theoretical side, multipartite entanglement may be a key element to improve various applications like quantum information processing or quantum metrology, or to understand and simulate physical systems, such as quantum spin chains undergoing a quantum phase transition. For multipartite systems, several measures of entanglement have been proposed. For example, generalized concurrence \cite{Carvalho14,Mintert16}, global entanglement \cite{Love18}, Scott measure (or generalized Meyer-Wallach measure) \cite{Meyer19,Brennen20,Scott21,Haddadi22}, geometric measures \cite{Qun23}, etc. Multipartite entanglement has been extensively investigated as a resource for quantum enhanced measurements. In the multipartite setting there are EMs that simply are functions of sums of bipartite EMs. For these multipartite EMs the monotonicity under LOCC is simply inherited from the bipartite measures. But there are also EMs which were constructed specifically for multipartite states. As required by an excellent or good EM, it can be checked that both bipartite and multipartite entanglement are two non-increasing EMs under LOCCs. Therefore it is natural to ask how much entanglement can be obtained from the incompletely entangled states which arise, for example, during the sharing of a completely entangled state between two observers using only LOCC. About fundamental questions of quantum information theory, of which tasks such as the characterization and general comprehension of entanglement belong to, LOCC operations are of importance because of their locality. As the concept of entanglement is strongly related to the nonlocal properties of a physical state \cite{Sheng24}, LOCC operations cannot affect the inherent nature of entanglement. By using LOCC operations different equivalence classes of states can be defined; representatives of each class can be used in experiments to carry out the same tasks, though with a different probability \cite{Nielsen10,Chitambar}.

\section{Conclusions}
\label{sec:4}
The amount of entanglement does not really have a meaning apart from a well-defined measure. An ideal measure of entanglement should have the following characteristics: it is non-vanishing if and only if the state is entangled; it is maximized by some recognizably maximally-entangled states; it has an operational interpretation (i.e., it quantifies the ability to carry out some quantum information protocol); it is monotonic (non-increasing under local operations and classical communication); and it is easy to calculate.  For bipartite pure states there is a measure that satisfies all of those requirements:  the entropy of entanglement, which is monotonic, straightforward to calculate, nonzero for all entangled states and zero for all product states, and which quantifies the number of maximally entangled pairs which can be produced asymptotically from many copies of the given state. However, for mixed states and multipartite states, there is no measure that satisfies all of these requirements as far as we know. There are variety of different measures that may satisfy some of these requirements but not others. Some (like negativity) are widely used in numerical modeling because they are easy to calculate, but in general they do not have a direct operational interpretation, and may not be nonzero for all entangled states.  Others have great theoretical importance (like the entanglement of formation) but cannot generally be calculated in closed form for most states. They require difficult optimizations, or regularized expressions, or both. Recently, Huang \cite{Huang} has proved that computing an EM is NP-hard if the measure is nonzero for all entangled states. Therefore, efficient algorithms or even closed analytical formulas for such EMs (including, but not limited to, entanglement of formation) are impossible unless P=NP.

\section*{Acknowledgments}
We would like to thank Yichen Huang, Yu-Bo Sheng, Ugo Marzolino, Todd A. Brun and Paul Erker for useful discussions. Saeed Haddadi would like to thank Marianne Bigornia and Somaye Ebrahimi for their helpful comments and the final edition of the paper.
%
%

\end{document}